\documentclass[aps,pra,twocolumn,amsmath,amssymb,showpacs]{revtex4}
\usepackage{amsmath}
\usepackage{graphicx}
\usepackage{epsfig}
\usepackage{amsfonts}

\begin{document}
\title{Parametric generation of quadrature squeezing of mirrors in cavity optomechanics}
\author{Jie-Qiao Liao}
\affiliation{Department of Physics and Institute of Theoretical
Physics, The Chinese University of Hong Kong, Shatin, Hong Kong
Special Administrative Region, People's Republic of China}
\author{C. K. Law}
\affiliation{Department of Physics and Institute of Theoretical
Physics, The Chinese University of Hong Kong, Shatin, Hong Kong
Special Administrative Region, People's Republic of China}
\date{\today}

\begin{abstract}
We propose a method to generate quadrature squeezed states of a
moving mirror in a Fabry-Perot cavity. This is achieved by
exploiting the fact that when the cavity is driven by an external
field with a large detuning, the moving mirror behaves as a
parametric oscillator. We show that parametric resonance can be
reached approximately by modulating the driving field amplitude at a
frequency matching the frequency shift of the mirror. The parametric
resonance leads to an efficient generation of squeezing, which is
limited by the thermal noise of the environment.
\end{abstract}

\maketitle

\section{Introduction}
Cavity
optomechanics~\cite{Review-Kippenberg,Review-Marquardt,Review-Karrai,Review-Aspelmeyer},
as an interaction interface between a cavity field and a moving
mirror, is an exciting research area for exploring quantum behavior
in macroscopic systems as well as applications in quantum
information processing. With the recent advances of cooling
techniques in optomechanical systems
~\cite{Wilson-Rae2007,Marquardt2007,Aspelmeyer2006,Heidmann2006,
Bouwmeester2006,Kippenberg2008,Gene2008,Wang2009}, it is becoming
possible to overcome thermal noise and study quantum state
engineering of mechanical mirrors. Indeed, recent studies have
already shown that various kinds of non-classical states can be
generated by optomechanical coupling.  These include quantum
superposition states~\cite{Bose1997,cat1}, entangled
states~\cite{Tombesi,Vedral2006,Vitali2007,Paternostro2007,Hartmann2008},
and squeezed states of light
~\cite{Fabre1994,Tombesi1994,Heidmann1994,Gong2009} and
mirrors~\cite{Ian2008,Zoller2009,Eisert2009,Wallquist2010,Girvin2010}.

Specifically, achieving squeezed states in mechanical oscillators
(mirrors) is an important goal because of the applications in
ultrahigh precision measurements such as the detection of
gravitational waves~\cite{Caves1980,Ligo1992,Weiss1999}. Several
schemes have been proposed to create quantum squeezing of the moving
mirror in cavity optomechanics. For example, squeezing can be
transferred from a squeezed light driving the cavity to the mirror
\cite{Zoller2009}, and recently Mari and Eisert have shown that
squeezing can be generated directly by a periodically modulated
driving field~\cite{Eisert2009}.

We note that a basic mechanism for creating quadrature squeezing is
to introduce a parametric coupling for the motional degree of
freedom of the mirror. In particular, efficient squeezing can be
achieved at the parametric resonance, such that the Hamiltonian in
the interaction picture takes the form $H_I  \propto b^2 + b^{\dag
2}$ [where $b$ and $b^\dag$ are operators of the oscillator in
Eq.~(\ref{Hamiltonian_S})] and the corresponding evolution operator
is a squeezed operator. Therefore an interesting question is how the
parametric resonance can be reached in cavity optomechanical
systems. One of the difficulties here is the dynamical shift of the
mechanical resonance frequency due to the optomehanical coupling,
which is sensitive to the intensity of the cavity field. In this
paper we show that in the large detuning limit, the frequency shift
can be compensated by modulating field amplitude at a suitable
frequency, and hence parametric resonance can be reached
approximately. We will present an explicit form of the driving
amplitude, and analyze the time development of squeezing in the
presence of thermal noise.

\section{Model}

\begin{figure}[ptb]
\center
\includegraphics[bb=44 648 277 756, width=3.3 in]{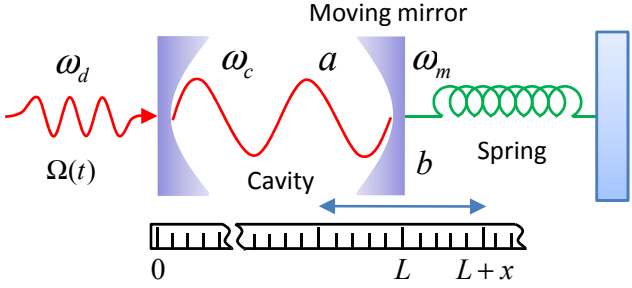}
\caption{(Color online) Schematic diagram of the cavity
optomechanical system. An externally driven Fabry-Perot cavity is
formed by a fixed end mirror and a harmonically bound end
mirror.}\label{setup}
\end{figure}

The system under consideration is an optical cavity formed by a
fixed mirror and a moving mirror connected with a spring
(Fig.~\ref{setup}). We consider a single-mode field in the cavity
and model the moving mirror as a harmonic oscillator. The
Hamiltonian of the system reads
\begin{eqnarray}
H_{S}&=&\hbar\omega_{c}a^{\dag}a+\hbar\omega_{m}
b^{\dagger}b-\hbar ga^{\dag}a(b^{\dagger}+b)\notag\\
&&+\hbar\Omega(t)e^{-i\omega_{d}t}a^{\dag}+\hbar\Omega^{\ast}(t)e^{i\omega_{d}t}a,
\label{Hamiltonian_S}
\end{eqnarray}
where $a^{\dag }$ ($b^{\dagger }$) and $a$ ($b$) are the creation
and annihilation operators associated with the single-mode cavity
field (mirror) with frequency $\omega_{c}$ ($\omega_m$). Assuming
$m_{\textrm{eff}}$ is the effective mass of the mirror, then the
position and momentum operators of the mirror are
$x=x_{\textrm{zpf}}(b^{\dag}+b)$ and $p=im_{\textrm{eff}}\omega
_{m}x_{\textrm{zpf}}({b}^{\dag}-b)$, where
$x_{\textrm{zpf}}=\sqrt{\hbar/(2m_{\textrm{eff}}\omega_{m})}$ is the
zero-point fluctuation of the mirror's position. The third term in
Eq.~(\ref{Hamiltonian_S}) describes a radiation pressure coupling
with the coupling strength $g=\omega_{c}x_{\textrm{zpf}}/L$, where
$L$ is the rest length of the cavity. In addition, the cavity is
driven by an external field with a main frequency $\omega_d$ and the
time-varying amplitude $\Omega(t)$.

To include damping in our model, we follow the standard approach by
coupling the system with oscillator baths such that the quantum
Langevin equations (in a rotating frame with frequency $\omega_{d}$)
for the operators $a$ and $b$ are given by
\begin{subequations}
\label{Langevineq}
\begin{align}
\dot{a}&=-i\Delta_{c}a+iga(b^{\dagger}+b)-i\Omega(t)-\frac{\gamma_{c}}{2}a+a_{in},\\
\dot{b}&=-i\omega_{m}b+iga^{\dag}a-\frac{\gamma_{m}}{2}b+b_{in},
\end{align}
\end{subequations}
with the detuning $\Delta_{c}=\omega_{c}-\omega_{d}$ and the cavity
(mirror) decay rate $\gamma_{c}$ ($\gamma_{m}$). Under the
assumption of Markovian baths, the noise operators $a_{in}$ and
$b_{in}$ have zero mean values and they are characterized by the
correlation functions $\langle
a_{in}(t)a^{\dagger}_{in}(t')\rangle=\gamma_{c}\delta(t-t')$,
$\langle a^{\dagger}_{in}(t)a_{in}(t')\rangle=0$, $\langle
b_{in}(t)b^{\dagger}_{in}(t')\rangle=\gamma_{m}(\bar{n}_{m}+1)\delta(t-t')$,
and $\langle
b^{\dagger}_{in}(t)b_{in}(t')\rangle=\gamma_{m}\bar{n}_{m}\delta(t-t')$,
where $\bar{n}_{m}=\{\exp[\hbar\omega_{m}/(k_{B}T_{m})]-1\}^{-1}$ is
thermal excitation number of the mirror's bath at temperature
$T_{m}$ and $k_{B}$ is the Boltzmann constant.  Here we have assumed
that the bath coupled to the cavity field is effectively a vacuum,
and the rotating-wave approximation has been employed to describe
the system-bath interaction~\cite{Dobrindt2008,Rodrigues2010}.

\section{Linearized system: Formal solution of fluctuations}

Next we write $a=\langle a \rangle  +\delta{a}$ and $b=\langle  b
\rangle +\delta{b}$ such that the fluctuations about the expectation
values are described by operators $\delta{a}$ and $\delta b$.
Assuming the fluctuations are sufficiently small, then we may
linearize Eq.~(\ref{Langevineq}) to obtain the equation of motion
for $\delta a$ and $\delta b$:
\begin{subequations}
\begin{align}
\delta\dot{a}&=-i\Delta(t)\delta a+ig\langle a(t)\rangle(\delta
b^{\dagger }+\delta b)-\frac{\gamma_{c}}{2}\delta
a+a_{in},\label{Leqda}\\
\delta \dot{b}&=-i\omega_{m}\delta b+ig[\langle
a^{\dagger}(t)\rangle\delta a+\langle a(t)\rangle \delta a^{\dag
}]-\frac{\gamma_{m}}{2}\delta b+b_{in},\label{Leqdb}
\end{align}
\end{subequations}
where $\Delta(t)=\Delta_{c}-g[\langle b(t)\rangle+\langle
b^{\dagger}(t)\rangle]$. The expectation values $\langle
a(t)\rangle$ and $\langle b(t)\rangle$ are governed by equations of
motion: $\dot{ \langle {a} \rangle}
=-[i\Delta(t)+\frac{\gamma_{c}}{2}]\langle {a} \rangle-i\Omega(t)$
and $\dot{\langle {b} \rangle} =-(i\omega_{m}+\frac{\gamma_{m}}{2})
\langle {b} \rangle +ig\vert \langle {a} \rangle \vert ^{2}$.

For convenience, we introduce the quadrature operators by $\delta
X_{s=a,b}=(\delta s^{\dag}+\delta s)/\sqrt{2}$ and $\delta
Y_{s=a,b}=i(\delta s^\dag - \delta s)/\sqrt{2}$. Then Eq. (3) can be
concisely expressed as
\begin{equation}
\dot{\mathbf{v}}(t)=\mathbf{M}(t)\mathbf{v}(t)+\mathbf{N}(t)\label{linearopoeq}
\end{equation}
where $\mathbf{v}=(\delta{X}_{a},\delta{Y}_{a},\delta{X}_{b},
\delta{Y}_{b})^{T}$, and ${\bf M}$ is
\begin{equation}
\mathbf{M}(t)=\left[
\begin{array}{cccc}
-\frac{\gamma _{c}}{2} & \Delta(t)& -\sqrt{2
}g\langle Y_{a}(t)\rangle & 0 \\
-\Delta(t)& -\frac{\gamma_{c}}{2} & \sqrt{2
}g\langle X_{a}(t)\rangle & 0 \\
0 & 0 & -\frac{\gamma _{m}}{2} & \omega _{m}\\
\sqrt{2}g\langle X_{a}(t)\rangle & \sqrt{2}g\langle Y_{a}(t)\rangle
& -\omega _{m} & -\frac{\gamma _{m}}{2}
\end{array}
\right],\label{Mmatrix}
\end{equation}
with $\langle X_{s=a,b}(t)\rangle=[\langle
s^{\dagger}(t)\rangle+\langle s(t)\rangle]/\sqrt{2}$ and $\langle
Y_{s=a,b}(t)\rangle=i[\langle s^{\dagger}(t)\rangle-\langle
s(t)\rangle]/\sqrt{2}$. The noise vector in Eq.~(\ref{linearopoeq})
is defined by $\mathbf{{N}}=({X}_{a}^{in}, {Y}_{a}^{in},
{X}_{b}^{in}, {Y}_{b}^{in})^{T}$, with
${X}_{s=a,b}^{in}=(s_{in}^{\dag }+s_{in})/\sqrt{2}$ and
${Y}_{s=a,b}^{in}=i(s_{in}^{\dag }-s_{in})/\sqrt{2}$.

Equation~(\ref{linearopoeq}) is a first-order linear inhomogeneous
differential equation with variable coefficients. Its formal
solution is
\begin{equation}
\mathbf{v}(t)=\mathbf{G}(t)\mathbf{v}(0)+\mathbf{G}(t)
\int_{0}^{t}\mathbf{G}^{-1}(\tau)\mathbf{{N}}(\tau)
d\tau,\label{formsoluofr}
\end{equation}
where the matrix $\mathbf{G}(t)$ satisfies $
\mathbf{\dot{G}}(t)=\mathbf{M}(t)\mathbf{G}(t)$ and the initial
condition $\mathbf{G}(0)=I$ ($I$ is the identity matrix). In the
present system, interesting quantities are the quadrature
fluctuations of the cavity and the mirror. Hence, we define a
covariance matrix $\mathbf{R}(t)$ by the elements
$\mathbf{R}_{ll'}(t)=\langle
\mathbf{v}_{l}(t)\mathbf{v}_{l'}(t)\rangle$ for $l,l'=1,2,3,4$.
Obviously, the four diagonal elements of $\mathbf{R}(t)$ are the
expectation values of the square of the four quadrature operators of
the system. They are
$\mathbf{R}_{11}(t)=\langle\delta{X}^{2}_{a}(t)\rangle$,
$\mathbf{R}_{22}(t)=\langle\delta{Y}^{2}_{a}(t)\rangle$,
$\mathbf{R}_{33}(t)=\langle\delta{X}^{2}_{b}(t)\rangle$, and
$\mathbf{R}_{44}(t)=\langle\delta{Y}^{2}_{b}(t)\rangle$. For the
mirror's rotating quadrature operator ${X}_{b}(\theta,t) \equiv
\cos\theta {X}_{b}(t)+\sin\theta {Y}_{b}(t)$,  the corresponding
variance is given by $\langle \delta
{X}_{b}^{2}(\theta,t)\rangle=\cos^{2}\theta \mathbf{R}_{33}(t)+\sin
^{2}\theta \mathbf{R}_{44}(t)+\frac{1}{2}\sin 2\theta
[\mathbf{R}_{34}(t)+\mathbf{R}_{43}(t)]$. Since
$[{X}_{b}(\theta,t),{X}_{b}(\theta+\pi/2,t)]=i$, quadrature
squeezing occurs when $\langle
\delta{X}_{b}^{2}(\theta,t)\rangle<1/2$.

To test the dynamical quadrature squeezing, we need to determine the
covariance matrix $\mathbf{R}(t)$, which has the formal expression:
\begin{equation}
\mathbf{R}(t)=\mathbf{G}(t)\mathbf{R}(0)
\mathbf{G}^{T}(t)+\mathbf{G}(t)\mathbf{Z}(t)\mathbf{G}^{T}(t).\label{solofR}
\end{equation}
where $\mathbf{Z}(t)$ is defined by
\begin{equation}
\mathbf{Z}(t)=\int_{0}^{t}\int_{0}^{t}\mathbf{G}^{-1}(\tau)
\mathbf{C}(\tau,\tau')[\mathbf{G}^{-1}(\tau')]^{T}d\tau d\tau' .
\label{defZmatrix}
\end{equation}
Here $\mathbf{C}(\tau,\tau')$ is the two-time noise operator
correlation matrix defined by the elements: $
\mathbf{C}_{nn'}(\tau,\tau')=\langle \mathbf{{N}}_{n}(\tau)
\mathbf{{N}}_{n^{\prime }}(\tau ^{\prime})\rangle$ for
$n,n'=1,2,3,4$. For Markovian baths, we have
$\mathbf{C}(\tau,\tau')=\mathbf{C}\delta(\tau-\tau^{\prime})$, where
the constant matrix $\mathbf{C}$ is given by
\begin{equation}
\mathbf{C}=\frac{1}{2}\left[
\begin{array}{cccc}
 \gamma_{c}& i\gamma_{c} & 0 & 0 \\
-i\gamma_{c} & \gamma_{c} & 0 & 0 \\
0 & 0 & \gamma_{m}(2\bar{n}_{m}+1) & i\gamma_{m} \\
0 & 0 & -i\gamma_{m} & \gamma_{m}(2\bar{n}_{m}+1)
\end{array}
\right].
\end{equation}

\section{Generation of quadrature squeezing}

Having obtained the formal equations for the evolution of quadrature
fluctuations of the mirror, we now ask how the external driving
amplitude $\Omega(t)$ can be chosen to generate a large degree of
quadrature squeezing of the mirror. We approach the problem by
considering the large detuning regime ($\Delta_{c}\gg\omega_{m}$) so
that by adiabatic approximation we have
\begin{equation}
\delta a\approx\frac{g} {\Delta_{c}-i \gamma_c/2} \langle
a(t)\rangle(\delta b^{\dagger}+\delta b)+F_{in},\label{adiabaticapp}
\end{equation}
with
$F_{in}=\int_{0}^{t}a_{in}(t')e^{(i\Delta_{c}+\gamma_{c}/2)(t'-t)}dt'$.
Here, we have also assumed $\Delta_{c}\gg g\langle X_{b}(t)\rangle$
and hence $\Delta(t)\approx\Delta_{c}$.  Correspondingly, the
equation of motion~(\ref{Leqdb}) for $\delta b$ becomes
\begin{equation}
\delta \dot{b}=-i\omega_{m}\delta b+i\eta |\langle a(t)\rangle|^2
(\delta b^{\dagger }+\delta b)-\frac{\gamma_{m}}{2}\delta
b+F'_{in},\label{effeqofdb}
\end{equation}
where $\eta = \frac{2g^2 \Delta_{c}} {\Delta_{c}^2 + \gamma_c^2/4}$
and the noise operator consists of two parts $F'_{in} \equiv
F^{a}_{in}+b_{in}$. The part $F^{a}_{in}=ig\langle
a^{\dagger}(t)\rangle F_{in}+ig\langle a(t)\rangle F^{\dagger}_{in}$
comes indirectly from the cavity's bath and depends on the mean
field solution, while the second part $b_{in}$ comes directly from
the mirror's bath.

Next we observe that if the external driving amplitude is chosen as
\begin{equation}
\Omega(t) = \Omega_0
\sin\left[\left(\omega_{m}-\xi_{0}\right)t\right]\label{Omegatform},
\end{equation}
with $\Omega_0$ being a constant and $\xi_{0}=g^{2}
\Omega_0^2\Delta_{c}/(\Delta_c ^2+\gamma_{c}^{2}/4)^{2}$, then by
the adiabatic solution $\langle
a(t)\rangle\approx-\Omega(t)/(\Delta_{c}-i\gamma_{c}/2)$ and the
assumption $\omega_{m}\gg \xi_{0}$, Eq.~(\ref{effeqofdb}) can be
approximated by
\begin{equation}
\delta  \dot{B}=-i\frac{\xi_{0}}{2} \delta B^{\dag}
-\frac{\gamma_{m}}{2}\delta
B+F'_{in}e^{i(\omega_{m}-\xi_{0})t},\label{eqofB}
\end{equation}
where $\delta B=\delta b e^{i(\omega_{m}-\xi_{0})t}$ is defined. In
deriving Eq.~(\ref{eqofB}), we have made use of a rotating wave
approximation such that counter-rotating terms with the rapidly
oscillating phase factors $e^{\pm 2i(\omega_{m}-\xi_{0})t}$ and
$e^{\pm 4i(\omega_{m}-\xi_{0})t}$ have been dropped.
\begin{figure}[ptb]
\center
\includegraphics[bb=33 7 457 312, width=3.3 in]{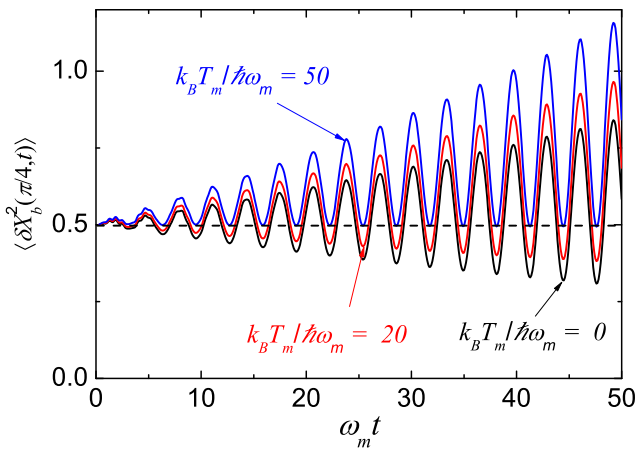}
\caption{(Color online) Time evolution of $\langle \delta
{X}_{b}^{2}(\pi/4,t)\rangle$ at various temperatures $T_{m}$. From
the bottom up, the three curves correspond to
$k_{B}T_{m}/(\hbar\omega_{m})=0,20$, and $50$, respectively. The
standard quantum limit $1/2$ is indicated by dashed black line. The
parameters are given in the text.} \label{dissipativecase}
\end{figure}

We notice that Eq.~(\ref{eqofB}) precisely corresponds to the
equation of motion of a damped parametric oscillator at resonance.
If damping can be ignored, a mirror initially prepared in the ground
state would display exponential squeezing as time increases:
$\langle \delta {X}_{b}^{2}(\pi/4,t)\rangle
=\frac{1}{2}e^{-\xi_{0}t}$. Such an efficient squeezing can be
understood by inspecting Eq.~(\ref{effeqofdb}) in which our choice
of $\Omega(t)$ matches the average value of the shifted resonance
frequency of the mirror $ \omega_m - \eta {|\langle
a(t)\rangle|}^2$, and therefore the parametric resonance can be
reached approximately. Note that $\xi_{0}$ is the average value of
such a frequency shift and it  also plays the role of an effective
strength of the parametric process.

However, it should be noted that for practical purposes, $\xi_0$ in
Eq.~(\ref{eqofB}), which decreases as $\Delta_c^{-3}$, has to be
strong enough to overcome noises of the baths, i.e., the detuning
$\Delta_c$ cannot be arbitrarily large. For realistic choices of
$\Delta_c$, the quality of squeezing has to be examined in the
presence of noise without making use of the adiabatic approximation.
To this end, we employ the linear formalism above and solve directly
the covariance matrix in Eq.~(\ref{solofR}) numerically. For
simplicity, we assume that the system is initially prepared in its
ground state $|0\rangle_{c}\otimes|0\rangle_{m}$ through a state
preparation process.  Such an initial state may be achievable in
future experiments based on ground-state cooling techniques. In
addition, we consider the following systems parameters:
$\omega_{m}=2\pi\times 1$~MHz, $\Delta_{c}=2\pi\times 10$~MHz,
$\gamma_{m}=2\pi\times 100$~Hz, $\gamma_{c}=2\pi\times 100$~kHz,
$\Omega_{0}\approx31.6$~GHz and $g=2\pi\times 100$~Hz, which are
realistic under current experimental
conditions~\cite{Aspelmeyer2009,Gorodetsky2010}. In
Fig.~\ref{dissipativecase} we plot the time-dependence of quadrature
variance of the mirror at various temperatures based on the form of
$\Omega (t)$ in Eq.~(\ref{Omegatform}), the evidence of squeezing is
clearly shown at sufficiently low temperatures. In fact, for not too
large $\Delta_c= 10 \omega_m$ chosen in Fig.~\ref{dissipativecase},
our exact numerical results agree well with the adiabatic
approximation.

If the temperature of the mirror's bath is higher than a critical
value, $T^{c}_{m}$, then there will no longer be squeezing in the
mirror (Fig.~\ref{dissipativecase}, blue line). A rough estimation
of the damping effect can be made by considering that the noise is
mainly from the mirror's bath, so that
\begin{equation}
\langle\delta X^{2}_{b}(\pi/4,t)\rangle \approx
\frac{1}{2}e^{-(\gamma_{m}+\xi_{0})t}+
\frac{\gamma_{m}(\bar{n}_{m}+1/2)}{(\gamma_{m}+\xi_{0})}\left(1-e^{-(\gamma_m+\xi_0)t}\right).
\end{equation}
Therefore squeezing occurs if the thermal excitation number
$\bar{n}_{m}$ is below a critical number, $\bar{n}^{c}_{m} =
\frac{\xi_{0}} { 2\gamma_{m}}$. For the parameters used in
Fig.~\ref{dissipativecase}, our estimation gives $T^{c}_{m}\approx
4.8$~mK or $k_{B}T^{c}_{m}/\hbar\omega_{m} \approx 50.5$, and this
agrees with the numerical value $50$ shown in
Fig.~\ref{dissipativecase}. We remark that in deriving Eq. (14), the
effect of noise $a_{in}$ has been neglected. This can be justified
by a lengthly calculation which shows that $\langle\delta X_b^{2}
\rangle $ due to the cavity field noise is of the order of
$\xi_0(\gamma_{m}+\xi_0+\gamma_{c})/[4\Delta_{c}(\gamma_{m}+\xi_0)]$
in the long time limit, and hence it can be made small compared with
the contribution from the thermal bath of the mirror by a large
detuning.

\section{Conclusion}

To conclude, we have presented a method to generate quadrature
squeezing of a mirror in cavity optomechanics. Specifically, we have
shown that in the large detuning regime with $\Delta_c \gg \omega_m
\gg \xi_0$ and $\Delta_c \gg g |\langle X_b (t) \rangle|$, the
driving field of the form $\Omega(t)$ given in Eq. (12) can generate
squeezing dynamically \cite{rmk}. The squeezing is supported by
direct numerical calculations for realistic parameters. We should
point out that our scheme is different from that in
Ref.~\cite{Eisert2009} because the large detuning regime considered
here enable us to eliminate the cavity field and formally map the
mirror to a parametric oscillator. In addition, parametric resonance
can be fine tuned by our driving field $\Omega(t)$ so that the
frequency shift of the mirror due to coupling to the cavity field
can be compensated approximately.

\begin{acknowledgments}
One of us (J.Q.L.) would like to thank Yu-Quan Ma and Nan Zhao for
technical support. This work is supported by the Research Grants
Council of Hong Kong, Special Administrative Region of China
(Project No.~CUHK401810).
\end{acknowledgments}


\begin{thebibliography}{99}
\bibitem{rmk}    The condition $\Delta_c \gg g |\langle X_b (t)
\rangle|$ is equivalent to $\omega_m \gg \xi_0 (1+ \gamma_c^2 /
\Delta_c^2)$ according to adiabatic approximation. For $\gamma_c <
\Delta_c$, the operation condition for our scheme is also $\Delta_c
\gg \omega_m \gg \xi_0$.
\end{thebibliography}
\end{document}